\documentclass[letterpaper, 10 pt, conference]{ieeeconf}

\IEEEoverridecommandlockouts
\usepackage[english]{babel}
\usepackage[utf8]{inputenc}
\usepackage[T1]{fontenc}

\usepackage{amsmath,amssymb,amsfonts,bm,mathtools}
\usepackage{graphicx}
\usepackage[font=small,labelfont=bf]{caption}
\usepackage[caption=false,font=footnotesize]{subfig}
\usepackage{booktabs}
\usepackage{url}
\usepackage[sort]{cite}
\usepackage{algorithm}
\usepackage{algorithmic}
\usepackage{xcolor}
\usepackage{soul}

\title{\LARGE \bf
Geometry-Informed Maritime Anomaly Detection Using Probabilistic Roadmaps
}

\author{Gabriele Oliva$^{*}$, Andrea Tomei, and Roberto Setola%
\thanks{Department of Engineering, Universit\`a Campus Bio-Medico di Roma, Via \'Alvaro del Portillo 21, 00128 Rome, Italy.}%
\thanks{$^{*}$Corresponding author. Email: g.oliva@unicampus.it}%
\thanks{This work was partly supported by project VIGIMARE, funded by the European Union under grant n. 101168016. Views and opinions expressed are however those of the authors only and do not necessarily reflect those of the European Union. Neither the European Union nor the granting authority can be held responsible for them.
This work was partly supported by Italian National project IMPROVE, funded by the Italian Ministry of Defense under grant n. 20711.}
}

\begin{document}
\maketitle
\thispagestyle{empty}
\pagestyle{empty}

\begin{abstract}
Maritime anomaly detection is essential for navigational safety and for the protection of critical underwater infrastructure. This paper proposes a geometry-informed supervised framework for detecting anomalous vessel trajectories in the Baltic Sea using Automatic Identification System (AIS) data. A Probabilistic Roadmap (PRM) is constructed over the navigable maritime domain and used as a structural prior to project trajectories onto feasible corridors. This representation enables the extraction of interpretable voyage-level features capturing route efficiency, geometric deviation from nominal paths, kinematic variability, and proximity to submarine cables. 
To address the scarcity of labeled anomalous events, synthetic anomalies are generated through controlled trajectory perturbations and infrastructure-aware distortions, producing a balanced dataset for supervised training. A Random Forest classifier is trained on the resulting feature set and evaluated under cross-validation and a held-out test split. Experimental results show stable generalization performance, achieving a test ROC AUC of 0.837, indicating the effectiveness of embedding navigational feasibility constraints into the anomaly detection process.
The proposed approach provides an interpretable and operationally relevant framework for infrastructure-aware maritime monitoring in geometrically complex environments.
\end{abstract}

\section{Introduction}

Maritime anomaly detection is a fundamental component of maritime domain awareness, supporting navigational safety, regulatory compliance, and the protection of critical infrastructure. The increasing availability of data from the Automatic Identification System (AIS) has enabled a new generation of data-driven monitoring frameworks, where vessel trajectories are analyzed to identify deviations from expected behavior~\cite{wolsing2022anomaly,ribeiro2023ais,yang2019big,dreo2022detection}. Recent developments increasingly rely on machine learning techniques capable of modeling complex spatiotemporal patterns in vessel motion~\cite{kim2024enhancing,xu2024sea}.

Early approaches focused on unsupervised modeling of nominal traffic patterns~\cite{vespe2012unsupervised,de2012machine}. Trajectory-based learning frameworks evolved toward clustering and probabilistic route modeling~\cite{pallotta2013vessel,zhen2017maritime}, followed by deep generative architectures such as Variational Recurrent Neural Networks and GeoTrackNet~\cite{nguyen2018multi,nguyen2021geotracknet}. More recently, hybrid graph-based and boosting-based approaches have been proposed for real-time or infrastructure-aware monitoring~\cite{guo2021anomaly,zaman2024online,mancini2024anomalous}. 

Despite their strong predictive performance, most existing methods rely primarily on latent behavior modeling or learned feature embeddings, without explicitly encoding navigational feasibility constraints arising from geography, coastlines, and restricted areas. As a result, geometric consistency is often inferred implicitly rather than structurally embedded in the anomaly definition.

The Baltic Sea represents a particularly challenging and strategically sensitive case study. It is characterized by dense traffic, archipelagos, narrow passages, and a high concentration of submarine communication and energy cables that are critical to European security and economic resilience~\cite{abels2025europe}. Recent infrastructure disruptions in the region have highlighted the vulnerability of underwater assets and the need for monitoring mechanisms capable of identifying subtle geometric deviations in vessel motion near critical infrastructure.

In this paper, we introduce a geometry-informed anomaly detection framework that integrates Probabilistic Roadmaps (PRMs)~\cite{kavraki1996probabilistic} with supervised learning. While PRMs were originally developed for robotic motion planning, we reinterpret them here as structural priors encoding feasible maritime corridors. AIS trajectories are projected onto this navigability graph, enabling the extraction of interpretable geometric, kinematic, and context-aware descriptors that quantify deviation from nominal routes and interaction with coastlines and submarine cables.

By combining PRM-informed feature engineering with supervised classification, the proposed framework balances interpretability and predictive performance, explicitly embedding spatial feasibility constraints into the anomaly definition and providing a structured alternative to purely latent trajectory representations.
\section{Proposed Framework}
\label{sec:framework}

This section presents the geometry-informed anomaly detection framework. 
We first describe the construction of the Probabilistic Roadmap (PRM) 
encoding navigational feasibility, then introduce the AIS dataset and 
scenario generation, and finally detail the trajectory projection 
mechanism that links observed vessel motion to the PRM structure.

\begin{figure}[t]
    \centering
    \includegraphics[width=0.45\textwidth]{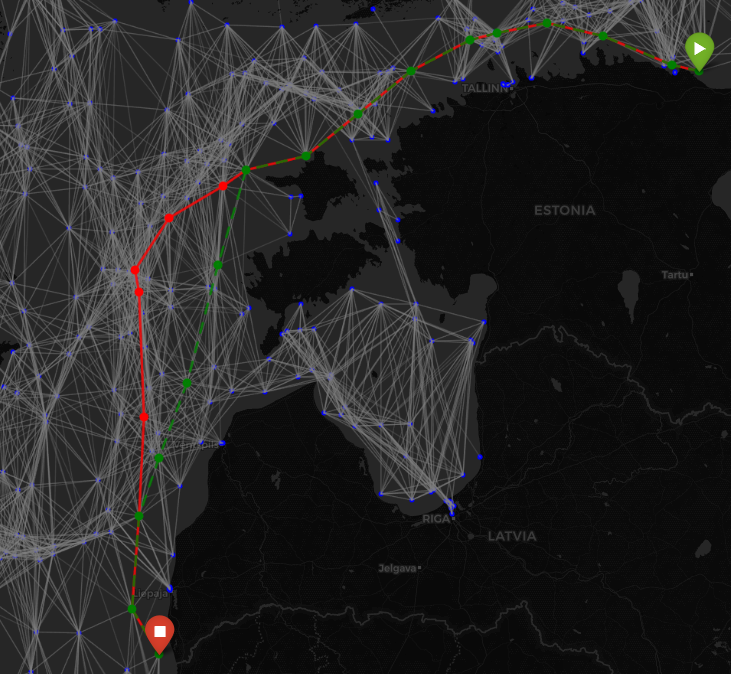}
    \caption{PRM-based route comparison: example of a non-economic path (red) versus the PRM shortest path (green).}
    \label{fig:noneconomic}
\end{figure}
\noindent {\bf Probabilistic Roadmap Construction.} A Probabilistic Roadmap (PRM)~\cite{kavraki1996probabilistic} 
is a sampling-based representation of a feasible configuration space, 
originally introduced in robotic motion planning. 
It consists of a weighted undirected graph $G = (V, E)$, where nodes \(V\) represent feasible configurations and edges \(E\) 
connect nearby configurations through collision-free paths.

In the maritime setting considered here, the configuration space 
corresponds to the navigable Baltic Sea domain. 
Feasible configurations are points located on water, 
while infeasible regions include land masses, coastlines, islands, 
and restricted areas. Two nodes are connected if the straight segment 
between them lies entirely within navigable water.

Each edge \(e \in E\) is assigned a weight equal to its Euclidean length 
in the projected coordinate reference system. 
For any two nodes \(v_i, v_j \in V\), the nominal maritime distance 
induced by the roadmap is defined as the shortest-path distance
\begin{equation}
d_G(v_i,v_j) = 
\min_{\pi \subseteq G} \sum_{e \in \pi} w(e),
\end{equation}
where \(\pi\) denotes a path in \(G\) and \(w(e)\) is the edge weight.

Unlike classical motion-planning applications, the PRM here is not used 
to compute control inputs but to encode a geometry-aware prior 
of navigational feasibility. It approximates large-scale maritime corridors 
and provides a structured baseline against which observed AIS trajectories 
can be compared.

Figure~\ref{fig:noneconomic} illustrates this principle. 
The red trajectory corresponds to a non-economic or deviating AIS path, 
while the green curve represents the shortest path computed on the PRM 
between the same endpoints. The discrepancy between the two provides 
a geometrically interpretable measure of route inefficiency or deviation 
from nominal navigational corridors.

To ensure connectivity while preserving sparsity, each node is linked 
to its \(k\)-nearest neighbors (kNN) in Euclidean distance. 
The node set is generated in two stages: the departure and arrival positions of the 858 real AIS voyages provide 1716 fixed nodes, and uniformly sampled points in the Baltic Sea bounding box bring the candidate set to 4000 nodes.
Candidate nodes are retained only if located 
on water using a land--water mask derived from Sentinel-2 imagery via 
Google Earth Engine~\cite{gorelick2017google,claverie2018harmonized}. 
Specifically, to confirm whether a point lies in the sea, a mask based on the reflectance of these images was implemented. Reflectance values range from 0 to 1, and a threshold of approximately 0.18 (an experimentally determined parameter) is used to classify a point as being in the sea, thereby qualifying it as a valid node. The advantage of using satellite data lies in its exceptional precision; for the B11 band, each pixel corresponds to a spatial resolution of 10-20 meters.

After filtering, 2284 valid nodes remain. We use \(k=25\) in EPSG:3035; preliminary sensitivity checks showed that smaller \(k\) fragmented narrow passages, while larger \(k\) mainly increased graph density. The reflectance threshold was visually checked against coastline layers.

Edges intersecting coastlines extracted from OpenStreetMap/Overpass 
are discarded~\cite{openstreetmap,overpassturbo}. 
The resulting graph supports efficient shortest-path queries 
\(d_G(\cdot,\cdot)\), used in subsequent feature extraction.

\noindent {\bf AIS Dataset and Coordinate Projection.} We consider general cargo (GC) vessel trajectories in the Baltic Sea 
between January~2017 and January~2019, obtained from the Finnish Marine Database~\cite{j3b5-es69-20}. 
Raw AIS messages are filtered to retain position reports and relevant static attributes.

Trajectories are cleaned by removing implausible records 
(e.g., unrealistic speeds or missing navigation status) 
and resampled to a uniform 5-minute interval. 
Routes shorter than 2 hours or covering less than 20~km are discarded, 
yielding 858 real trajectories.

AIS positions are originally expressed in geographic coordinates 
(WGS84, EPSG:4326). Since Euclidean distance computations are required 
for PRM queries and trajectory metrics, all coordinates are projected 
into the European Lambert Azimuthal Equal-Area system (EPSG:3035), 
which provides consistent metric distances over the Baltic region.

To incorporate spatial awareness, we also integrate coastline data 
from OpenStreetMap via Overpass~\cite{openstreetmap,overpassturbo} 
and submarine cable polylines from TeleGeography~\cite{submarinecablemap}. 
These layers enable infrastructure-aware feature extraction 
and interaction analysis with critical underwater assets.

\noindent {\bf Synthetic Anomaly Generation.} Because labeled anomalous trajectories are scarce in real AIS data, 
we generate 858 synthetic anomalies to obtain a balanced dataset 
(1716 total trajectories). The synthetic generation process aims to 
introduce controlled yet structurally plausible deviations from nominal 
routes while preserving physical and navigational consistency. 
All generated trajectories retain the original departure and arrival points.

The anomalous samples are evenly distributed among the following mechanisms:

\begin{enumerate}

\item \textbf{Sinusoidal Deviation.}  
This method introduces systematic lateral perturbations to nominal AIS trajectories 
through a periodic model. Let the original latitude–longitude sequence be 
\((\phi(t), \lambda(t))\). The perturbed trajectory is defined as \mbox{$\phi'(t) = \phi(t) + A \sin(\omega t)$}, \mbox{$\lambda'(t) = \lambda(t) + A \cos(\omega t)$}, where \(A\) controls the amplitude and \(\omega\) the frequency of deviation. 
To simulate realistic positioning uncertainty, Gaussian noise proportional to \(A\) 
is added to both coordinates. Headings are recomputed using geodetic bearing formulas 
to maintain directional coherence. In our experiments, 
\(A = 0.0015\) degrees and \(\omega = 0.05\) were selected to produce moderate 
yet detectable deviations.

\item \textbf{Markov-Based Perturbations.}  
A probabilistic deviation model is constructed using a 
\(3\times3\) directional probability kernel

\[
P =
\begin{bmatrix}
p_{NW} & p_{N} & p_{NE} \\
p_{W}  & p_{C} & p_{E}  \\
p_{SW} & p_{S} & p_{SE}
\end{bmatrix},
\]

where \(p_X\) denotes the probability of moving in direction \(X\) 
(and \(p_C\) the probability of remaining stationary), 
with \(\sum_X p_X = 1\). 
Here \(P\) is a local categorical distribution, not a grid-state transition matrix: its nine entries are flattened, normalized, and sampled to choose one of eight compass moves or the stay action. Route endpoints are then reinserted for comparison with the original voyage.
To induce anomalous behavior, we generate matrices with approximately uniform 
probabilities but introduce a mild bias toward the vessel’s current heading. 
At each step, a direction is sampled according to \(P\), and latitude–longitude 
increments are computed with trigonometric corrections accounting for 
Earth curvature. Endpoints are preserved while intermediate segments are 
stochastically modified.

\item \textbf{Autoencoder-Based Distortions.}  
An unsupervised autoencoder~\cite{bank2023autoencoders} 
is trained to learn a compact latent representation of nominal trajectories. 
After training, controlled Gaussian noise (standard deviation \(0.10\)) 
is injected into the normalized input before reconstruction. 
The decoded output produces structurally coherent yet statistically altered 
routes. Headings are recomputed to preserve kinematic plausibility. 
This approach generates nonlinear distortions that differ from parametric 
sinusoidal or Markov perturbations.

\item \textbf{Speed Anomalies Near Submarine Cables.}  
To introduce infrastructure-aware deviations, we identify trajectory segments 
intersecting a 500-meter buffer around submarine cable polylines. 
For a randomly selected subset of trajectories, vessel speed along these segments 
is multiplied by a factor between 1.8 and 2.2. 
This simulates abnormal maneuvers near critical underwater assets, 
directly interacting with the contextual features used in classification.
\end{enumerate}

The anomalous dataset is evenly distributed across generation mechanisms 
(approximately 286 trajectories per method) to avoid overfitting to a single 
perturbation type. By combining deterministic, stochastic, generative, 
and infrastructure-aware distortions, the synthetic set spans a diverse 
spectrum of plausible non-nominal maritime behaviors while maintaining 
basic physical consistency.

\noindent {\bf PRM-Based Trajectory Projection.} Let a trajectory be \(p_1,\dots,p_N\) in the projected CRS. 
Each AIS point is associated to the closest PRM node:
\begin{equation}
\hat v_i = \arg\min_{v\in V}\|p_i-v\|_2.
\end{equation}

Define the geometric trajectory length
\begin{equation}
L_{\text{AIS}}=\sum_{i=1}^{N-1}\|p_{i+1}-p_i\|_2,
\end{equation}
and the PRM-induced length
\begin{equation}
L_{\text{PRM}}=\sum_{i=1}^{N-1} d_G(\hat v_i,\hat v_{i+1}),
\end{equation}
where \(d_G(\cdot,\cdot)\) denotes shortest-path distance on the PRM graph.

Quantities such as the ratio \(L_{\text{AIS}}/L_{\text{PRM}}\), 
pointwise deviations \(\|p_i-\hat v_i\|_2\), and residence time near 
repeated PRM nodes provide interpretable indicators of 
deviation from nominal maritime corridors.

\section{Feature Extraction}
\label{sec:features}

We extract voyage-level descriptors grouped into three categories:
(i) \emph{PRM conformity}, 
(ii) \emph{kinematics}, and 
(iii) \emph{contextual interaction}. 
Let a trajectory be \( \{p_i\}_{i=1}^N \), with sampling interval \(\Delta t\), projected onto PRM nodes \(\{\hat v_i\}\) as described above. The geometric trajectory length is denoted \(L_{\text{AIS}}\), while \(L_{\text{PRM}}\) denotes the corresponding shortest-path length on the PRM graph.

\noindent {\bf PRM conformity features.} We quantify deviation from nominal navigation corridors using:

\begin{align}
\texttt{length-ratio} &= \frac{L_{\text{AIS}}}{L_{\text{PRM}}},\\
\texttt{length-diff}  &= L_{\text{AIS}} - L_{\text{PRM}},\\
\texttt{max-dev} &= \max_{i}\|p_i-\hat v_i\|_2,\\
\texttt{avg-dev} &= \frac{1}{N}\sum_{i=1}^N \|p_i-\hat v_i\|_2.
\end{align}

The \texttt{length-ratio} measures route efficiency relative to the PRM baseline: values close to one indicate conformity with established corridors, while larger values suggest detours or non-economic paths. The \texttt{length-diff} captures the absolute excess distance traveled. The quantities \texttt{max-dev} and \texttt{avg-dev} represent, respectively, the maximum and mean Euclidean deviation between trajectory points and their closest PRM nodes.

We further compute PRM coverage,
\begin{equation}
\texttt{PRM-coverage}=\frac{1}{N}\left|\left\{i:\|p_i-\hat v_i\|_2<\omega\right\}\right|,
\end{equation}
where \(\omega\) is a tolerance threshold. This feature measures the proportion of the trajectory lying close to nominal corridors. Low coverage indicates persistent off-route navigation.

To detect loitering behavior, we also evaluate the maximum cumulative time associated with a single projected PRM node, which captures prolonged stationary or near-stationary activity.

\noindent {\bf Kinematic features}

Let \(v_i\) and \(\psi_i\) denote the speed and heading at point \(p_i\), and let \(\bar v\), \(\bar \psi\) denote their voyage-level averages.

\begin{align}
\texttt{speed-var} &= \frac{1}{N-1}\sum_{i=1}^N (v_i-\bar v)^2,\\
\texttt{max-abs-acc} &= \max_{i\ge2}\left|\frac{v_i-v_{i-1}}{\Delta t}\right|,\\
\texttt{heading-var} &= \frac{1}{N-1}\sum_{i=1}^N(\psi_i-\bar\psi)^2.
\end{align}

The feature \texttt{speed-var} captures variability in cruising behavior; elevated values may indicate erratic motion. The \texttt{max-abs-acc} measures the largest instantaneous acceleration or deceleration magnitude, potentially revealing abrupt maneuvers. The \texttt{heading-var} quantifies directional instability, with large values corresponding to frequent or sharp turns.

\noindent {\bf Context features (coastlines and cables).} To encode spatial awareness, we compute the minimum distance to coastline geometries:
\begin{equation}
\texttt{coast-dist}=\min_{i}\ \mathrm{dist}(p_i,\mathcal{C}),
\end{equation}
where \(\mathcal{C}\) denotes the coastline set. Smaller values indicate navigation in close proximity to shore, which may be atypical depending on vessel class and route.

For submarine cables \(\mathcal{A}\), we define a buffer region \(B(\mathcal{A},d)\) of radius \(d\) (e.g., 100--500\,m) and compute interaction metrics:

\begin{align}
\iota_i(d) &= \mathbf{1}_{\{[p_i,p_{i+1}]\cap B(\mathcal{A},d)\neq\emptyset\}},\\
\texttt{time-near-cable} &= \sum_{i=1}^{N-1}\Delta t\ \iota_i(d),\\
\texttt{speed-near-cable} &= 
\frac{\sum_{i=1}^{N-1} v_{i+1}\ \iota_i(d)}
{\sum_{i=1}^{N-1}\iota_i(d)}.
\end{align}

Here, \(\iota_i(d)\) is the indicator function, equal to 1 when the trajectory segment intersects the cable buffer and 0 otherwise. The feature \texttt{time-near-cable} measures the cumulative duration spent near critical infrastructure, while \texttt{speed-near-cable} captures the vessel’s average speed during such interactions. Unusual lingering or abnormal speed patterns near cables may signal suspicious activity.
\section{Classification}
\label{sec:classification}

Anomaly detection is formulated as a supervised binary classification problem at the voyage level. 
Given the feature vector \(x \in \mathbb{R}^d\) extracted as described in Section~\ref{sec:features}, the goal is to learn a mapping
\[
f : \mathbb{R}^d \rightarrow \{0,1\},
\]
where label \(1\) denotes an anomalous trajectory and \(0\) a normal one.

In this work, we adopt a Random Forest (RF) classifier~\cite{ho1995random}, a tree-based ensemble method particularly suitable for heterogeneous feature sets combining geometric, kinematic, and contextual descriptors.

\noindent {\bf Random Forest Model.} Random Forest constructs an ensemble of \(B\) decision trees \(\{T_b\}_{b=1}^B\), each trained on a bootstrap sample \(\mathcal{D}_b\) drawn with replacement from the training set. 
At each split of a tree, a random subset of \(m\) features (with \(m \le d\)) is considered, introducing additional decorrelation between trees and improving generalization.

For binary classification, the ensemble prediction is obtained by majority voting \mbox{$\hat{y}_{RF}(x) = 
\mathrm{mode}\left\{T_b(x)\right\}_{b=1}^B,$} where \(T_b(x) \in \{0,1\}\) denotes the class predicted by the \(b\)-th tree.

The anomaly probability estimate is computed as the average of per-tree probabilities:
\begin{equation}
P_{RF}(y=1 \mid x) = \frac{1}{B} \sum_{b=1}^B P_b(y=1 \mid x).
\end{equation}

Each tree is grown by recursively partitioning the feature space so as to maximize impurity reduction. 
We adopt the Gini impurity criterion \mbox{$G(S) = 1 - \sum_{c \in \{0,1\}} p_c^2$}, where \(p_c\) denotes the proportion of class \(c\) samples in node \(S\). 
At each split, the feature and threshold are selected to maximize the impurity decrease.

\noindent {\bf Model Selection and Evaluation.} Hyperparameters such as the number of trees \(B\), maximum depth, minimum samples per split, minimum samples per leaf, and maximum number of features per split are tuned via cross-validation using a two-stage strategy: an initial random search over a broad parameter range, followed by a local grid refinement around the best configuration.

Model performance is evaluated using accuracy, precision, recall, F1-score, and ROC AUC. 
The ROC AUC metric, defined as \mbox{$\text{ROC AUC} = \int_0^1 \text{TPR}(t)\, d\text{FPR}(t),$} quantifies the classifier’s ability to rank anomalous trajectories above normal ones across all decision thresholds \(t\).

Random Forest is particularly well-suited to the present setting because it:
(i) handles nonlinear interactions between heterogeneous features,
(ii) is robust to moderate feature correlations,
(iii) provides intrinsic measures of feature importance, and
(iv) offers strong generalization performance without requiring feature scaling assumptions.

\section{Numerical Results}
\label{sec:numericalresults}

The proposed anomaly detection framework was evaluated on a balanced dataset comprising 858 real general cargo trajectories from the Baltic Sea and 858 synthetically generated anomalous trajectories. The data were split into training (70\%), validation (10\%), and test (20\%) subsets to ensure an unbiased performance assessment.
The split is performed at voyage level, hyperparameter search uses only training folds, and no anomaly-generator identifier or perturbation parameter is used as an input feature. The PRM is kept fixed as a label-free geographic prior.

\noindent {\bf Cross-Validation and Model Selection.} Random Forest hyperparameters were optimized using the two-stage search strategy described in Section~\ref{sec:classification}. Model selection was based on 10-fold cross-validation performed on the training set.

Across validation folds, the classifier achieved a mean accuracy of 73.2\%, precision of 74.7\%, recall of 70.2\%, F1-score of 72.3\%, and ROC AUC of 0.813. These values indicate a balanced trade-off between false positives and false negatives, with stable ranking capability across thresholds. 

For reference, training performance reached 86.5\% accuracy and an ROC AUC of 0.949. The gap between training and validation metrics suggests moderate overfitting, yet the degradation remains limited and consistent with the heterogeneous and high-dimensional nature of the feature space.
These results should be read as a controlled stress test: the cable-speed perturbation is intentionally aligned with one contextual feature, while the other perturbation families are not. Independent real anomalies are still needed for external validation.

\begin{figure*}[htbp]
    \centering
    \includegraphics[width=0.75\textwidth]{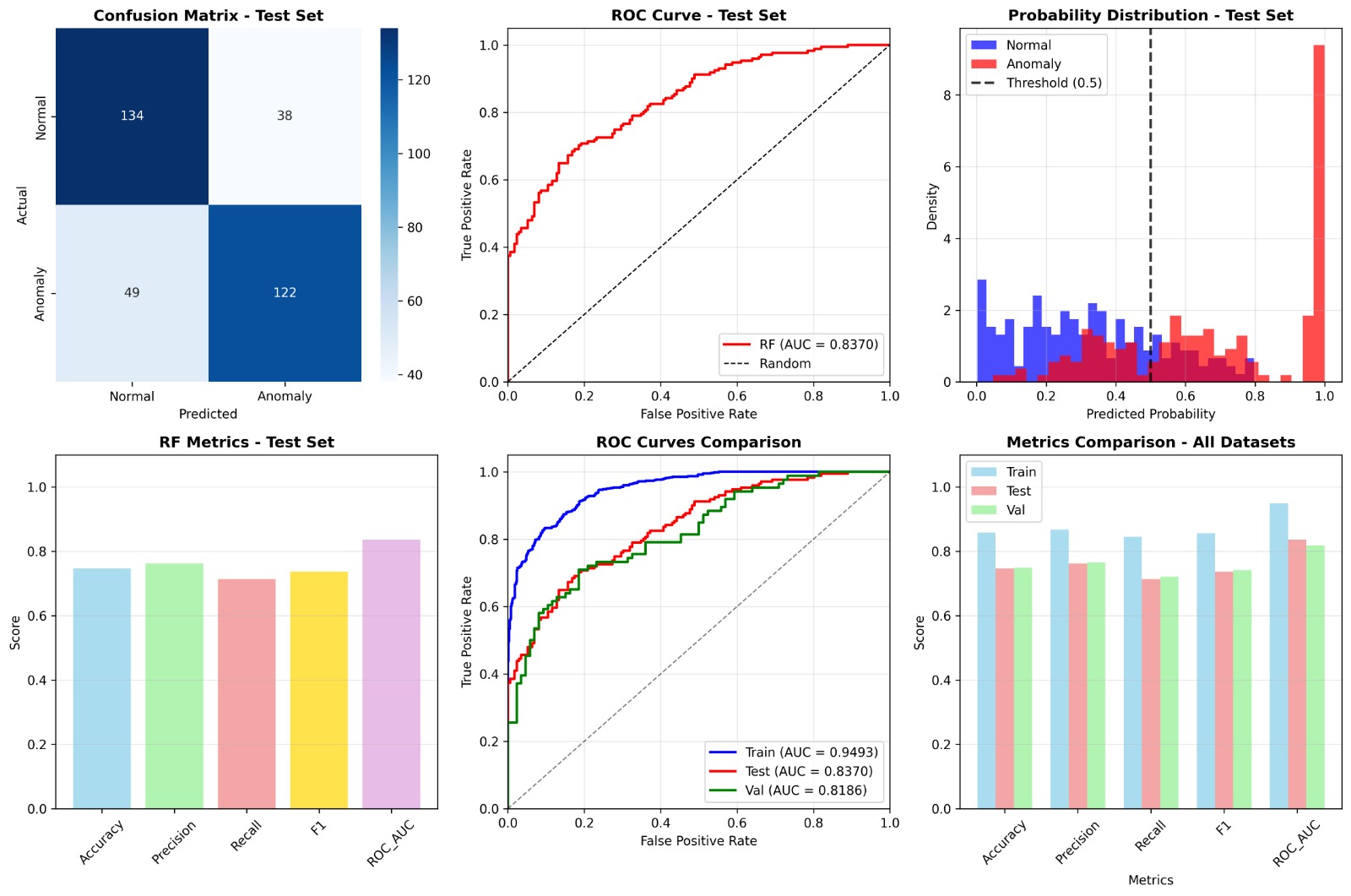}
    \caption{Random Forest performance analysis: (a) confusion matrix, (b) ROC curve on the test set, (c) probability density distribution for normal and anomalous classes with decision threshold at 0.5, (d) performance metrics for the test set, (e) cross-validation ROC AUC across folds, and (f) summary performance metrics.}
    \label{fig:ensembleanalysis}
\end{figure*}

\noindent {\bf Test Set Evaluation.} Final evaluation on the held-out test set is summarized in Fig.~\ref{fig:ensembleanalysis}. 
The confusion matrix (Fig.~\ref{fig:ensembleanalysis}a) reports 134 true negatives and 122 true positives, with 38 false positives and 49 false negatives. This corresponds to a test accuracy of 74.6\%, precision of 76.2\%, recall of 71.3\%, F1-score of 73.7\%, and ROC AUC of 0.837. The consolidated test metrics are also shown in Fig.~\ref{fig:ensembleanalysis}d.

The ROC curve (Fig.~\ref{fig:ensembleanalysis}b) confirms strong discriminative capability, with clear separation from the random baseline. Importantly, the test AUC (0.837) remains close to the validation AUC (0.819), while the training AUC (0.949) is higher but does not indicate severe overfitting. The variability of ROC AUC across cross-validation folds (Fig.~\ref{fig:ensembleanalysis}e) remains moderate, suggesting that performance is not driven by a specific data partition and is consistent with the combined PRM, kinematic, and contextual representation.

The distribution of predicted anomaly probabilities (Fig.~\ref{fig:ensembleanalysis}c) shows that anomalous trajectories concentrate toward higher probability values, whereas normal trajectories are predominantly located below the decision threshold $\tau=0.5$. Although partial overlap remains in the intermediate region, the separation is sufficiently structured to support threshold tuning in operational contexts. A summary comparison of performance metrics across training, validation, and test sets is provided in Fig.~\ref{fig:ensembleanalysis}f, further illustrating consistent generalization behavior.

Overall, the results indicate that embedding navigational feasibility constraints via the PRM yields a structurally meaningful representation of maritime behavior, enabling robust anomaly discrimination in a geometrically complex environment.
The absence of kinematic-only and non-PRM geometric baselines remains a limitation; these ablations are needed to isolate the roadmap contribution.

\section{Conclusions and Perspectives}
\label{sec:conclusions}

This paper presented a geometry-informed framework for maritime anomaly detection based on Probabilistic Roadmaps (PRMs) and supervised learning. 
By encoding navigational feasibility constraints into a structured graph representation of the Baltic Sea domain, the proposed approach embeds spatial consistency directly into the feature space. 
Projecting AIS trajectories onto the PRM enables the extraction of interpretable voyage-level descriptors capturing route efficiency, geometric deviation from nominal corridors, kinematic variability, and interaction with critical underwater infrastructure.

Experimental results on real and synthetically augmented AIS data demonstrate stable generalization performance, with a test ROC AUC of 0.837. 
These findings indicate that incorporating geometric priors into the anomaly definition can provide an interpretable complement to latent or trajectory-only representations, although a broader baseline comparison is required to quantify the gain.

Several extensions naturally emerge from this framework. 
First, the PRM representation can be exploited beyond feature extraction by defining anomaly scores directly at the graph level, for instance through path-consistency measures or probabilistic route likelihood models. 
Second, incorporating temporal dynamics such as seasonal traffic patterns or density-dependent priors would allow the PRM to evolve from a static navigability graph into a time-aware maritime behavior model. 
Third, integrating uncertainty quantification in both trajectory projection and classification would provide calibrated risk estimates suitable for operational deployment.
Environmental and weather information, such as wind, currents, storms, bathymetry, and traffic-separation schemes, can be added as time-varying PRM costs or voyage-level covariates. Future experiments will also include feature-group ablations and sensitivity analyses for node density, \(k\), water-mask threshold, and cable-buffer radius.

Overall, the proposed framework establishes a structured and infrastructure-aware basis for maritime anomaly detection, paving the way toward probabilistic and dynamic graph-based modeling in future work.

\addtolength{\textheight}{-1.10in}
\bibliographystyle{IEEEtran}
\bibliography{references}

@article{kim2024enhancing,
  title={Enhancing generalization of active sonar classification using semisupervised anomaly detection with multisphere for normal data},
  author={Kim, Geunhwan and Choo, Youngmin},
  journal={IEEE Journal of Oceanic Engineering},
  volume={49},
  number={4},
  pages={1530--1548},
  year={2024},
  publisher={IEEE}
}

@article{xu2024sea,
  title={Sea Surface Floating Small Target Detection Based on a Priori Feature Distribution and Multiscan Iteration},
  author={Xu, Shuwen and Zhang, Tian and Ru, Hongtao},
  journal={IEEE Journal of Oceanic Engineering},
  year={2024},
  publisher={IEEE}
}

@article{claverie2018harmonized,
  title={The Harmonized Landsat and Sentinel-2 surface reflectance data set},
  author={Claverie, Martin and Ju, Junchang and Masek, Jeffrey G and Dungan, Jennifer L and Vermote, Eric F and Roger, Jean-Claude and Skakun, Sergii V and Justice, Christopher},
  journal={Remote sensing of environment},
  volume={219},
  pages={145--161},
  year={2018},
  publisher={Elsevier}
}

@article{gorelick2017google,
  title={Google Earth Engine: Planetary-scale geospatial analysis for everyone},
  author={Gorelick, Noel and Hancher, Matt and Dixon, Mike and Ilyushchenko, Simon and Thau, David and Moore, Rebecca},
  journal={Remote sensing of Environment},
  volume={202},
  pages={18--27},
  year={2017},
  publisher={Elsevier}
}

@article{bank2023autoencoders,
  title={Autoencoders},
  author={Bank, Dor and Koenigstein, Noam and Giryes, Raja},
  journal={Machine learning for data science handbook: data mining and knowledge discovery handbook},
  pages={353--374},
  year={2023},
  publisher={Springer}
}

@misc{openstreetmap,
  author       = {{OpenStreetMap contributors}},
  title        = {OpenStreetMap},
  year         = {2025},
  howpublished = {\url{https://www.openstreetmap.org}},
  note         = {Accessed: 2025-05-24}
}

@misc{overpassturbo,
  author       = {Martin Raifer},
  title        = {Overpass Turbo},
  year         = {2025},
  howpublished = {\url{https://overpass-turbo.eu}},
  note         = {Accessed: 2025-05-24}
}

@misc{submarinecablemap,
  author       = {TeleGeography},
  title        = {Submarine Cable Map},
  year         = {2025},
  howpublished = {\url{https://www.submarinecablemap.com}},
  note         = {Accessed: 2025-05-24}
}

@data{j3b5-es69-20,
doi = {10.21227/j3b5-es69},
url = {https://dx.doi.org/10.21227/j3b5-es69},
author = {Ville Hakola},
publisher = {IEEE Dataport},
title = {Vessel tracking (AIS), vessel metadata and dirway datasets},
year = {2020} }

@article{kavraki1996probabilistic,
  title={Probabilistic Roadmaps for Path Planning in High-Dimensional Configuration Spaces},
  author={Kavraki, Lydia E. and Svestka, P. and Latombe, Jean-Claude and Overmars, Mark H.},
  journal={IEEE Transactions on Robotics and Automation},
  year={1996},
  volume={12},
  number={4},
  pages={566--580}
}

@article{abels2025europe,
  title={EUROPE AND THE SECOND COLD WAR IN SUBMARINE CABLE NETWORKS},
  author={ABELS, Joscha},
  year={2025}
}

@inproceedings{ho1995random,
  title={Random decision forests},
  author={Ho, Tin Kam},
  booktitle={Proceedings of 3rd international conference on document analysis and recognition},
  volume={1},
  pages={278--282},
  year={1995},
  organization={IEEE}
}

@article{yang2019big,
  title={How big data enriches maritime research--a critical review of Automatic Identification System (AIS) data applications},
  author={Yang, Dong and Wu, Lingxiao and Wang, Shuaian and Jia, Haiying and Li, Kevin X},
  journal={Transport reviews},
  volume={39},
  number={6},
  pages={755--773},
  year={2019},
  publisher={Taylor \& Francis}
}

@article{dreo2022detection,
  title={Detection and localization of multiple ships using acoustic vector sensors on buoyancy gliders: Practical design considerations and experimental verifications},
  author={Dreo, Richard and Trabattoni, Alister and Stinco, Pietro and Micheli, Michele and Tesei, Alessandra},
  journal={IEEE Journal of Oceanic Engineering},
  volume={48},
  number={2},
  pages={577--591},
  year={2022},
  publisher={IEEE}
}

@article{wolsing2022anomaly,
  title={Anomaly detection in maritime AIS tracks: A review of recent approaches},
  author={Wolsing, Konrad and Roepert, Linus and Bauer, Jan and Wehrle, Klaus},
  journal={Journal of Marine Science and Engineering},
  volume={10},
  number={1},
  pages={112},
  year={2022},
  publisher={MDPI}
}

@article{ribeiro2023ais,
  title={AIS-based maritime anomaly traffic detection: A review},
  author={Ribeiro, Claudio V and Paes, Aline and de Oliveira, Daniel},
  journal={Expert Systems with Applications},
  volume={231},
  pages={120561},
  year={2023},
  publisher={Elsevier}
}

@article{pallotta2013vessel,
  title={Vessel pattern knowledge discovery from AIS data: A framework for anomaly detection and route prediction},
  author={Pallotta, Giuliana and Vespe, Michele and Bryan, Karna},
  journal={Entropy},
  volume={15},
  number={6},
  pages={2218--2245},
  year={2013},
  publisher={MDPI}
}

@article{guo2021anomaly,
  title={An anomaly detection method for AIS trajectory based on kinematic interpolation},
  author={Guo, Shaoqing and Mou, Junmin and Chen, Linying and Chen, Pengfei},
  journal={Journal of Marine Science and Engineering},
  volume={9},
  number={6},
  pages={609},
  year={2021},
  publisher={Multidisciplinary Digital Publishing Institute}
}

@article{zaman2024online,
  title={Online Ornstein--Uhlenbeck based anomaly detection and behavior classification using AIS data in maritime},
  author={Zaman, Bakht and Marijan, Dusica and Kholodna, Tetyana},
  journal={Ocean Engineering},
  volume={312},
  pages={119057},
  year={2024},
  publisher={Elsevier}
}

@article{nguyen2021geotracknet,
  title={GeoTrackNet—A maritime anomaly detector using probabilistic neural network representation of AIS tracks and a contrario detection},
  author={Nguyen, Duong and Vadaine, Rodolphe and Hajduch, Guillaume and Garello, Ren{\'e} and Fablet, Ronan},
  journal={IEEE Transactions on Intelligent Transportation Systems},
  volume={23},
  number={6},
  pages={5655--5667},
  year={2021},
  publisher={IEEE}
}

@inproceedings{nguyen2018multi,
  title={A multi-task deep learning architecture for maritime surveillance using AIS data streams},
  author={Nguyen, Duong and Vadaine, Rodolphe and Hajduch, Guillaume and Garello, Ren{\'e} and Fablet, Ronan},
  booktitle={2018 IEEE 5th International Conference on Data Science and Advanced Analytics (DSAA)},
  pages={331--340},
  year={2018},
  organization={IEEE}
}

@article{zhen2017maritime,
  title={Maritime anomaly detection within coastal waters based on vessel trajectory clustering and Na{\"\i}ve Bayes Classifier},
  author={Zhen, Rong and Jin, Yongxing and Hu, Qinyou and Shao, Zheping and Nikitakos, Nikitas},
  journal={The Journal of Navigation},
  volume={70},
  number={3},
  pages={648--670},
  year={2017},
  publisher={Cambridge University Press}
}

@article{de2012machine,
  title={Machine learning for vessel trajectories using compression, alignments and domain knowledge},
  author={De Vries, Gerben Klaas Dirk and Van Someren, Maarten},
  journal={Expert Systems with Applications},
  volume={39},
  number={18},
  pages={13426--13439},
  year={2012},
  publisher={Elsevier}
}

@inproceedings{vespe2012unsupervised,
  title={Unsupervised learning of maritime traffic patterns for anomaly detection},
  author={Vespe, Michele and Visentini, Ingrid and Bryan, Karna and Braca, Paolo},
  booktitle={9th IET Data Fusion \& Target Tracking Conference (DF\&TT 2012): Algorithms \& Applications},
  pages={14--1},
  year={2012},
  organization={IET}
}

@inproceedings{mancini2024anomalous,
  title={Anomalous Vessel Behavior Detection via Offline Clustering of Regular Trajectories},
  author={Mancini, Giulia Di Berto and Fioravanti, Camilla and Sacchetti, Marco and Oliva, Gabriele and Setola, Roberto},
  booktitle={2024 32nd Mediterranean Conference on Control and Automation (MED)},
  pages={730--735},
  year={2024},
  organization={IEEE}
}
\end{document}